\documentclass[twocolumn,showpacs,preprintnumbers,amsmath,amssymb,prl,unsortedaddress,superscriptaddress]{revtex4-1}
\usepackage[pdftex]{graphicx}
\usepackage{dcolumn}         
\usepackage{bm}              

\begin{document}

\title{Electronic Structure of Liquid Water and a Platinum Surface}

\author{Isaac Tamblyn}
  \email{isaac.tamblyn@uoit.ca}
  \affiliation{Department of Physics, University of Ontario Institute of Technology, Oshawa, ON, L1H 7K4, Canada}
  
\begin{abstract}

Many-body perturbation theory within the G$_0$W$_0$ approximation is used to determine molecular orbital level alignment at a liquid water/Pt(111) interface generated through \emph{ab initio} molecular dynamics. Molecular orbital energy levels are shown to depend both on the position of H$_2$O molecules within the liquid relative to the surface and the details of their local bonding environment. Standard density functional theory calculations disagree qualitatively with level alignment predicted by many-body perturbation theory.
\end{abstract}

\pacs{73.30.+y, 79.60.Jv, 71.15.Mb, 31.15.A-} 

\maketitle


Platinum surfaces are well known for their superior catalytic properties, enabling reactions ranging from the reduction of CO$_2$~\cite{white_jco2_2014} into fuels to high efficiently water-splitting~\cite{michaelides_jacs_2001}. Unfortunately, platinum is not an Earth abundant element, precluding its cost effective deployment as a grid-scale electrocatalyst. Identifying alternatives to Pt would have significant consequences for applications in short term energy storage~\cite{diaz_rser_2012} and carbon neutral fuel cycles. Developing an \emph{ab initio}, microscopic understanding of this ideal interface will aid efforts to identify cheaper, more abundant alternatives~\cite{cheng_pccp_2012, willard_jcp_2013}.

Density functional theory (DFT) is now the standard method used to study nano-scale structures, including liquids, bulk materials, and interfaces ~\cite{becke_jcp_2014}. Unfortunately, it has the well known limitation of inaccurate predictions of frontier orbital energies~\cite{kummel_rmp_2008}. This is not surprising, as formally only the energy of the highest occupied molecular orbital (HOMO) must be correct (it corresponds to the 1$^{\textrm{st}}$ vertical ionization energy, IE)~\cite{PPLB1982, Almbladh1985}. Typical approximations to the exchange-correlation potential lack the correct long range, ~-1/r asymptotic behavior (crucial for describing electron addition/removal processes)~\cite{kummel_rmp_2008}, posses no derivative discontinuity\cite{Perdew1983}, and suffer from self-interaction errors. As a result, ionization energies are consistently underestimated, particularly in the case of low molecular weight species~\cite{tamblyn_jpcl_2014}. Interfaces present an additional challenge for DFT~\cite{pulci_prl_1998, rignanese_prl_2001, shaltaf_prl_2008}, as molecular levels can be strongly affected by static and dynamical polarization ~\cite{neaton_prl_2006, garcia-lastra_prb_2009, thygesen_prl_2009, freysoldt_prl_2009, chen_jcp_2010}- both forms of non-local correlation which are not well described by typical functionals. The mismatch of self-interaction errors between small molecules (with highly localized electrons) and the itinerant electrons present in metals make such environments particularly susceptible to erroneous bonding, hybridization, and charge transfer. 

\begin{figure}[!t]
\begin{center}
\includegraphics[width=0.4\textwidth,clip]{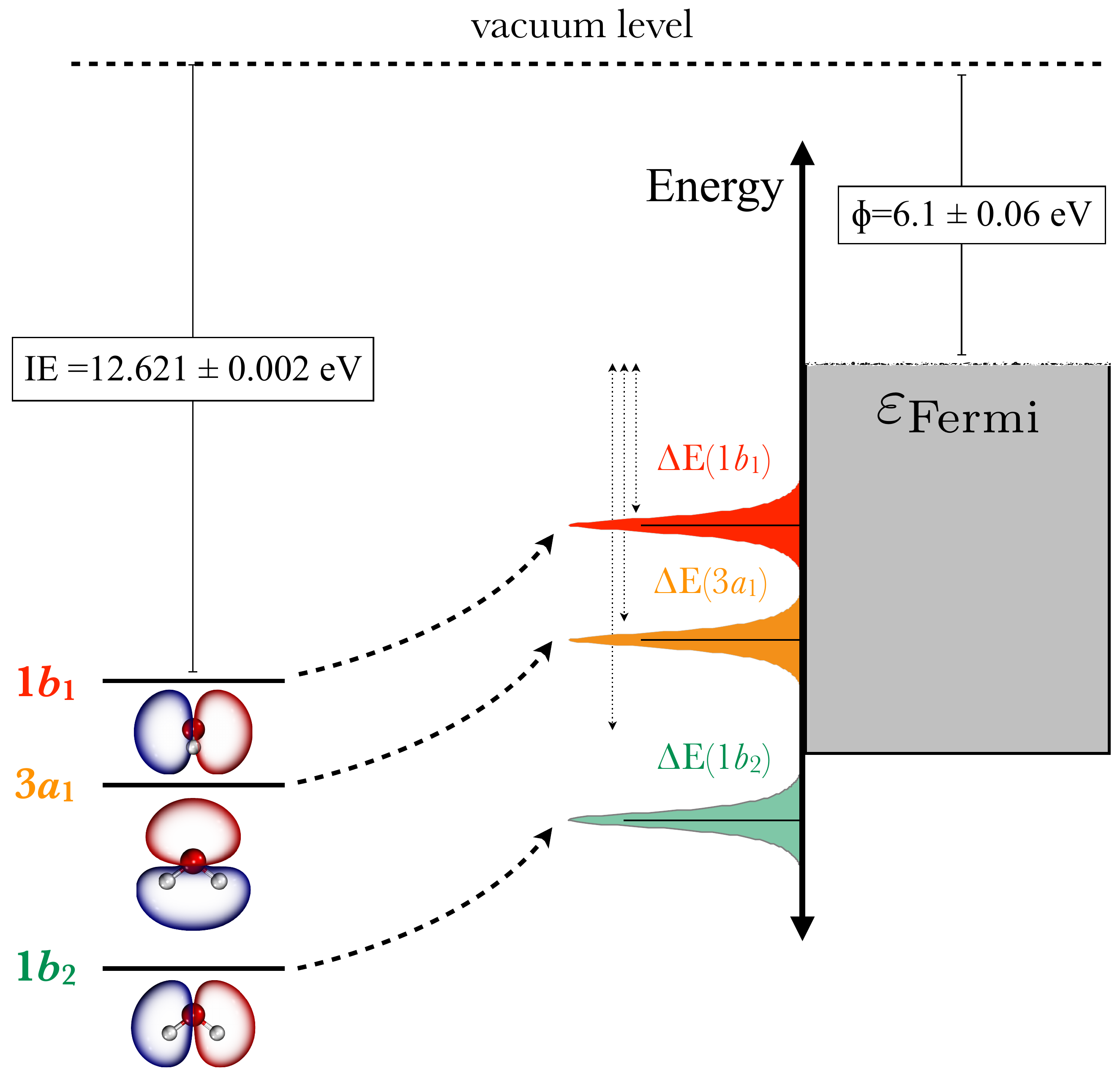}
\end{center}
\caption{\label{f:level_sketch} Experimental ionization energy and work function  for gas phase water~\cite{faubel_jcp_1997, NIST} and the Pt(111) surface~\cite{smith_prb_1974, shevchik_prb_1976, derry_prb_1989}. When a liquid is brought into contact with a surface, the resulting energy level alignment ($\Delta E$) depends on these energies, modified by the self-consistent charge transfer and rearrangement (and corresponding induced dipole), hybridization, and substrate polarization.}
\end{figure}


In recent years, a powerful perturbative method has seen wider use within in the electronic structure community. The GW approximation expands the self-energy in terms of the single-particle Green's function~\cite{martin_pr_1959} and the screened Coulomb interaction~\cite{hedin_physrev_1965}, $W=\epsilon^{-1}V$, (where $V$ is the ``bare'' Coulomb interaction and $\epsilon$ is the energy dependant dielectric function). The self-energy operator, $\hat\Sigma$, is an energy dependant operator with eigenvectors corresponding to addition and removal quasi-particle states (and corresponding energy). The GW approach has now been successfully applied to bulk metals~\cite{northrup_prb_1989, faleev_prb_2010}, semiconductors~\cite{hybertsen_prb_1986}, isolated molecules~\cite{rostgaard_prb_2010, blase_prb_2011}, and liquids~\cite{garbuio_prl_2006, pham_prb_2014}. 

With increased interest in electrolysis and photocatalytic water splitting for applications in renewable energy, it is likely that in coming years, the number of DFT calculations on such systems will increase substantially. It is therefore timely to provide an accurate assessment of this important interface using a method capable of describing the multitude of physical effects which determine level alignment.

\begin{figure}[!t]
\begin{center}
\includegraphics[width=0.5\textwidth,clip]{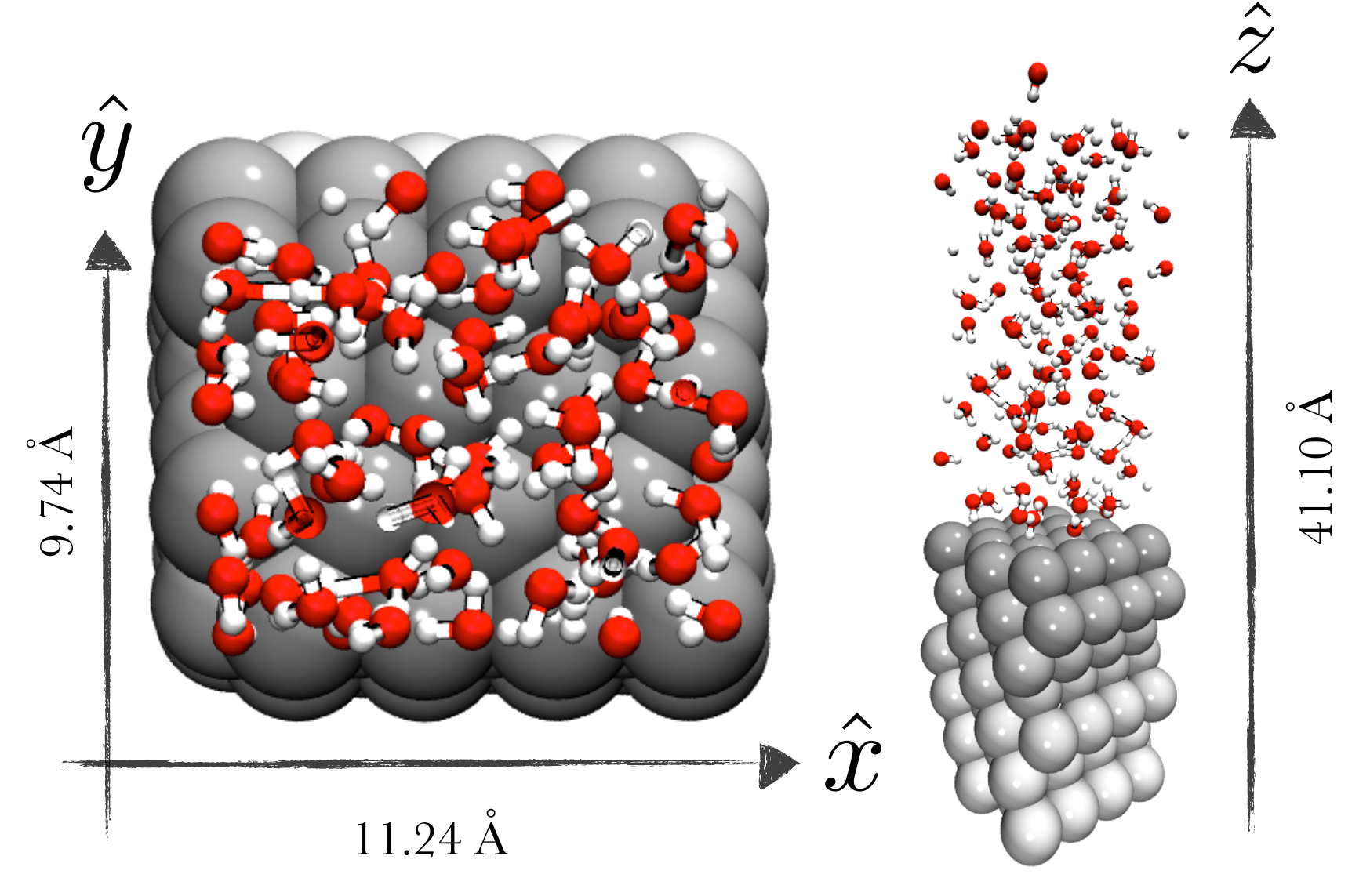}
\end{center}
\caption{\label{f:geometry} Geometry of the liquid water/Pt(111) interface used in \emph{ab initio} molecular dynamics and G$_0$W$_0$ level alignment calculations. The density at the midpoint of the (periodic) slabs is 1 g/cc.}
\end{figure}

Obtaining an accurate description of liquid water with \emph{ab initio} methods is challenging. Recent work~\cite{mcmahon_arxiv_2014, morales_jctc_2014} suggests that both quantum nuclear and non-local (i.e. vdW) interactions play important roles determining the structure of water at ambient conditions. Applying such methods to a system of the size studied here is currently intractable, and the correct treatment of transition metals with vdW-capable DFT is still an open question~\cite{li_prb_2012}. Earlier work with water\cite{schwegler_jcp_2004, vandeVondele_jcp_2005, todorova_jpcb_2006, sharma_prl_2007, cicero_jacs_2008, morrone_prl_2008} noted that simulations at elevated temperatures (T=330-400 K) tended to cancel errors in methods which lacked the two aforementioned physics. For these reasons, I have elected to use an established (DFT-PBE\cite{pbe}) method to handle the geometry (using the ``trick'' of higher temperature). To generate an metal-water interfacial geometry (Fig.~\ref{f:geometry}), I first relaxed a periodic slab (4$\times$7$\times$7 layers with a (111) surface) in the absence of the liquid, holding the $z$-component of the inner 3 metal layers fixed at the theoretical bulk lattice constant (3.975~\AA). Next, I placed 100 deuterium molecules in an amorphous geometry~\cite{packmol} in the vacuum region. The length of the cell in the $z$ direction was chosen so that the average density in the bulk fluctuates around 1 g/cc. After a partial geometry relaxation~\cite{vasp}, I equilibrated the system with a Nose-Hoover thermostat (T=330K) for 3 ps (integration timestep was 0.27 fs) and randomly selected a snapshot for the subsequent G$_0$W$_0$ calculation.  I used a plane-wave basis and Brillouin Zone sampling (2$\times$2$\times$1) with a $\Gamma$-point centred mesh.

For the G$_0$W$_0$ results, I included 18 electrons in the valence shell of the pseudopotential, rather than the typical (for DFT calculations) approach of placing 5s and 5p in the core~\cite{marini_prl_2001}. Given the 100 water molecules also present, this results in a large number (2800+) of electrons to consider. I used a 60 Ry cutoff in the planewave expansion of the Kohn-Sham states. 

In the implementation used here (BerkeleyGW~\cite{BerkeleyGW}), evaluation of both $\epsilon$ and $\hat\Sigma$ involve a summation over unoccupied states. To converge my calculations, I used 2680 empty states. This corresponds to an integration window of $\approx$ 35 eV beyond the Fermi level ($\varepsilon_\textrm{Fermi}$). To represent $\epsilon$, I used cutoff of 20 Ry. Smaller values can result in false convergence behaviour~\cite{sharifzadeh_epjb_2012}. The energy dependence of $\epsilon$ was treated with the Hybertsen-Louie~\cite{hybertsen_prb_1986} plasmon-pole model. Achieving absolute convergence (of level position) in a GW calculation is difficult, and several closure techniques now exist. I elected to use the static remainder method~\cite{deslippe_prb_2013} due to its proven performance predicting ionization energies of small molecules~\cite{sharifzadeh_epjb_2012}. Based on the convergence parameters discussed above, for an isolated water molecule, the G$_0$W$_0$ ionization energy is 12.0 eV (5\% underestimate of experiment). As previously discussed~\cite{tamblyn_prb_2011}, subsystem errors (molecular IE and metal work function) carry over to the interface. Thus the level alignment reported here likely has a similar offset. To achieve an IE within 1\% of experiment, I find that the isolated molecule requires a 60 Ry cutoff for $\epsilon$ and integration over an energy window 90 eV above $\varepsilon_\textrm{Fermi}$. Given that the calculations reported here already require a substantial computing platform, (32 TB RAM), achieving 1\% accuracy using this approach is likely to be to outside the realm of ``routine calculations'' for several years.

In heterogeneous systems, where the DFT eigenvalues for the composite pieces suffer from varying magnitudes of errors (i.e. delocalized metal states are well described within typical DFT functionals whereas localized molecular ones are not), mixing which occurs during the DFT self-consistency cycle can be very different than what would have occurred in the absence of such errors. Thus quasi-particle states could be quite different than the Kohn-Sham ones; $\hat\Sigma$ is unlikely to be diagonal in this basis~\cite{pulci_prb_1999, rangel_prb_2011}. One is then faced with two options - evaluate $\hat\Sigma$ for all KS$_{ij}$ pair (\emph{and as a function of energy}) or find a more suitable basis in which the diagonal approximation is more appropriate.

Within the unit cell, I label the molecules according to their (z) position, $i=1...N_m$, (here $N_m=100$). $i=1$ is the molecule closest (and likely adsorbed) to the left-hand wall and $i=N_m$ is next to the right hand wall. For each molecule's instantaneous geometry, there exist a set of ``gas-phase''~\cite{tamblyn_prb_2011, yu_jpcl_2013} orbitals $\{| \iota_i \rangle\}$ which are eigenstates of the Hamiltonian, $\hat H^{i}$. For the left-most molecule, they satisfy: $\hat H^{i=1} | \iota^{i=1} \rangle = \epsilon^{i=1} | \iota^{i=1} \rangle$. The superscript in $\hat H$ denotes which (single) molecular geometry (in the gas phase) was used to define the Hamiltonian.

\begin{figure}[!h]
\begin{center}
\includegraphics[width=0.5\textwidth,clip]{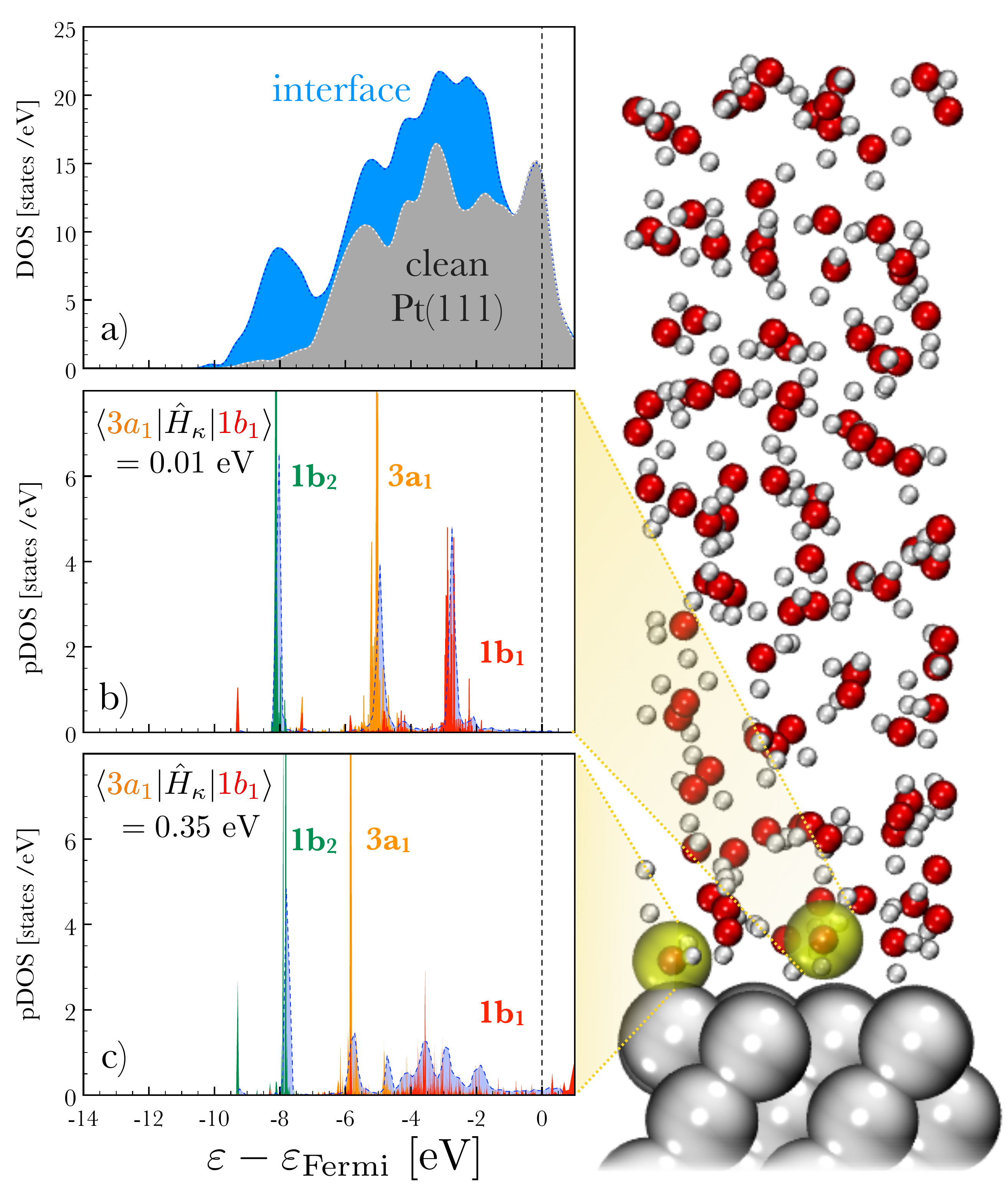}
\end{center}
\caption{\label{f:matrix} a) Electronic density of states for clean Pt(111) and the liquid/Pt interface. b) Projected density of states for a molecule ($i=3$, see text) in the liquid close to the Pt surface. c) Projected density of states for a molecule ($i=1$) adsorbed to the surface. By projecting onto gas phase orbitals it is possible to identify level position and hybridization of orbitals associated with different molecules. A projection onto the atomic oxygen p orbitals (blue filled curve) is also included for comparison. Depending on the local bonding environment and interaction with the surface, the degree of hybridization of molecular orbitals varies significantly. This can be seen in the onsite, off-diagonal coupling between gas phase orbitals.}
\end{figure}

In this notation, $\langle \iota_{\textrm{1b1}}^{i=1} | \hat H^{i=1} | \iota_{\textrm{1b1}}^{i=1} \rangle$ means the (HOMO) energy of a gas phase molecule with the \emph{geometry} of the molecule adsorbed to the left-hand wall. Since each molecule has a different geometry (due to the environment from which they were extracted) $\langle \iota_{\textrm{1b1}}^{i=1} | \hat H^{i=1} | \iota_{\textrm{1b1}}^{i=1} \rangle \neq \langle \iota_{\textrm{1b1}}^{i=2} | \hat H^{i=2} | \iota_{\textrm{1b1}}^{i=2} \rangle \neq ... \neq \langle \iota_{\textrm{1b1}}^{i=Nm} | \hat H^{i=Nm} | \iota_{\textrm{1b1}}^{i=Nm} \rangle$ (though they are of similar magnitude). Beyond evaluating $\hat \Sigma$, these orbitals form a useful set with which to construct a projected density of states (Fig.~\ref{f:matrix}), and can be used to interpret the electronic structure of the interface. They compare well with the complimentary approach of projecting directly onto atomic states (oxygen p states in this case). The significant broadening (and off-diagonal coupling) observed for the surface H$_2$O molecule is an indicator of bonding.

\begin{figure}[!t]
\begin{center}
\includegraphics[width=0.5\textwidth,clip]{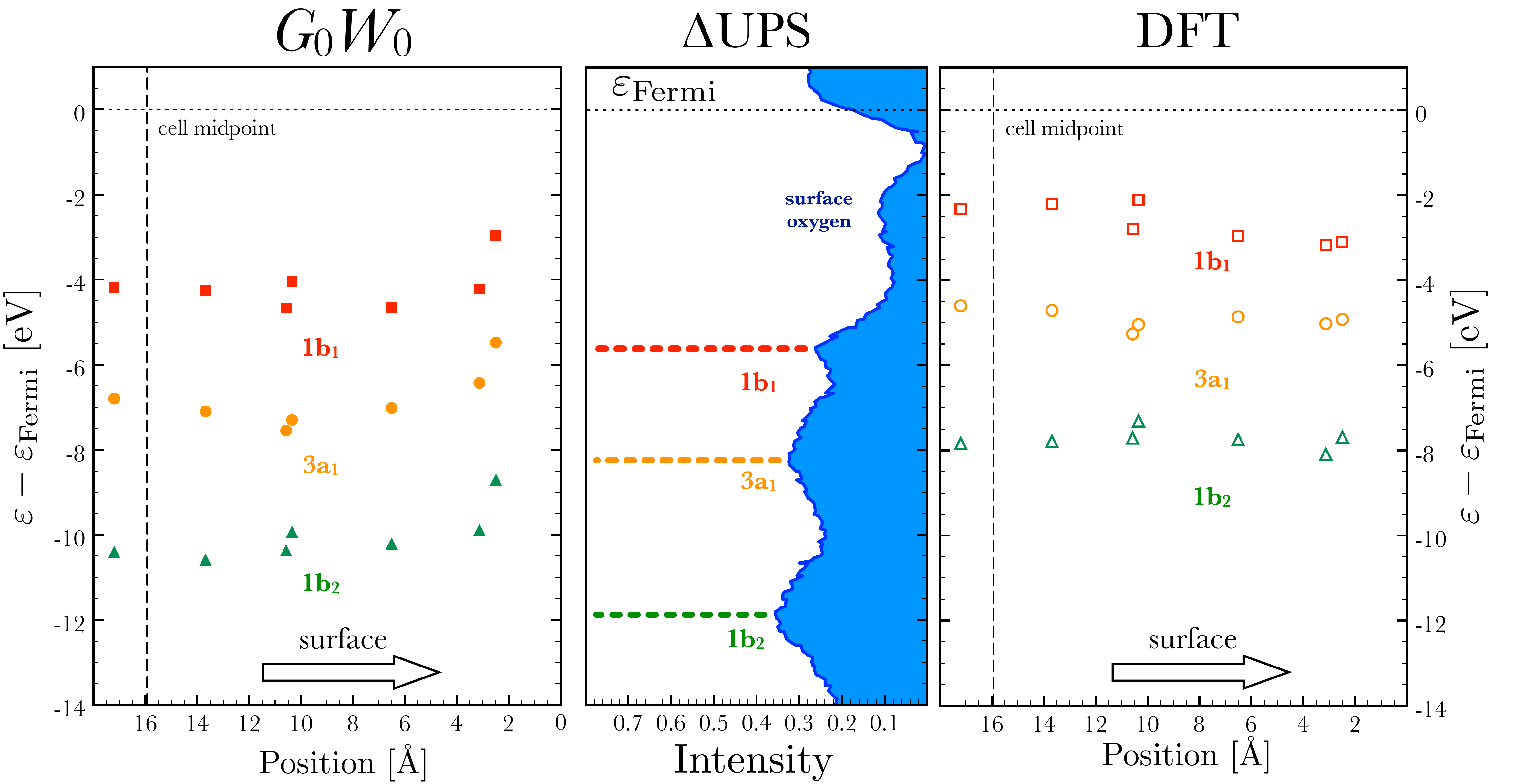}
\end{center}
\caption{\label{f:gw} Orbital level alignment ($\Delta E$ from Fig.~\ref{f:level_sketch} and defined in the text) as a function of position relative to the metal surface within a) G$_0$W$_0$, b) UPS, and c) DFT. The experimental signal, b), is based on the difference in measurements of clean and hydrated Pt(111). In G$_0$W$_0$, orbitals near the surface are more shallow (energetically) than those in the bulk. DFT exhibits the opposite trend (see also Fig. ~\ref{f:qp_shift}a) with respect to position.}
\end{figure}

We are interested in the orbital energies of molecules in the presence of the liquid and surface. Thus the relevant Hamiltonian is that of the combined system, $H^{\kappa}$. The energy offsets noted in Fig.~\ref{f:level_sketch} are given by $\Delta E(1b1)^{i} = \langle \iota_{\textrm{1b1}}^{i} | \hat H^{\kappa} | \iota_{\textrm{1b1}}^{i} \rangle - \varepsilon_{\textrm{Fermi}}$. $\Delta E(3a1)^i$ and $\Delta E(3a1)^i$ are defined similarly. In Fig.~\ref{f:gw}, I plot $\Delta E(1b1)$, $\Delta E(3a1)$, and $\Delta E(1b2)$ at the level of G$_0$W$_0$ and DFT for molecules at various distances from the metal surface, along with experimental UPS measurements.

Interestingly, for the case of a molecule bound to the surface (as shown in Fig.~\ref{f:matrix}c), G$_0$W$_0$ and DFT give essentially the same value. Both sit at an energy which corresponds (experimentally) to bound oxygen (peak assignment is based on the fact that oxide layers on Pt(111) and H$_2$O/Pt(111) both have signal here). The fact that DFT and G$_0$W$_0$ give similar values (the quasi-particle shift is small) for this species is consistent with the fact that orbitals which are well hybridized with a surface tend to suffer from self-interaction errors of similar magnitude to those of the substrate itself - this results in a form of error cancellation.

\begin{figure}[!t]
\begin{center}
\includegraphics[width=0.4\textwidth,clip]{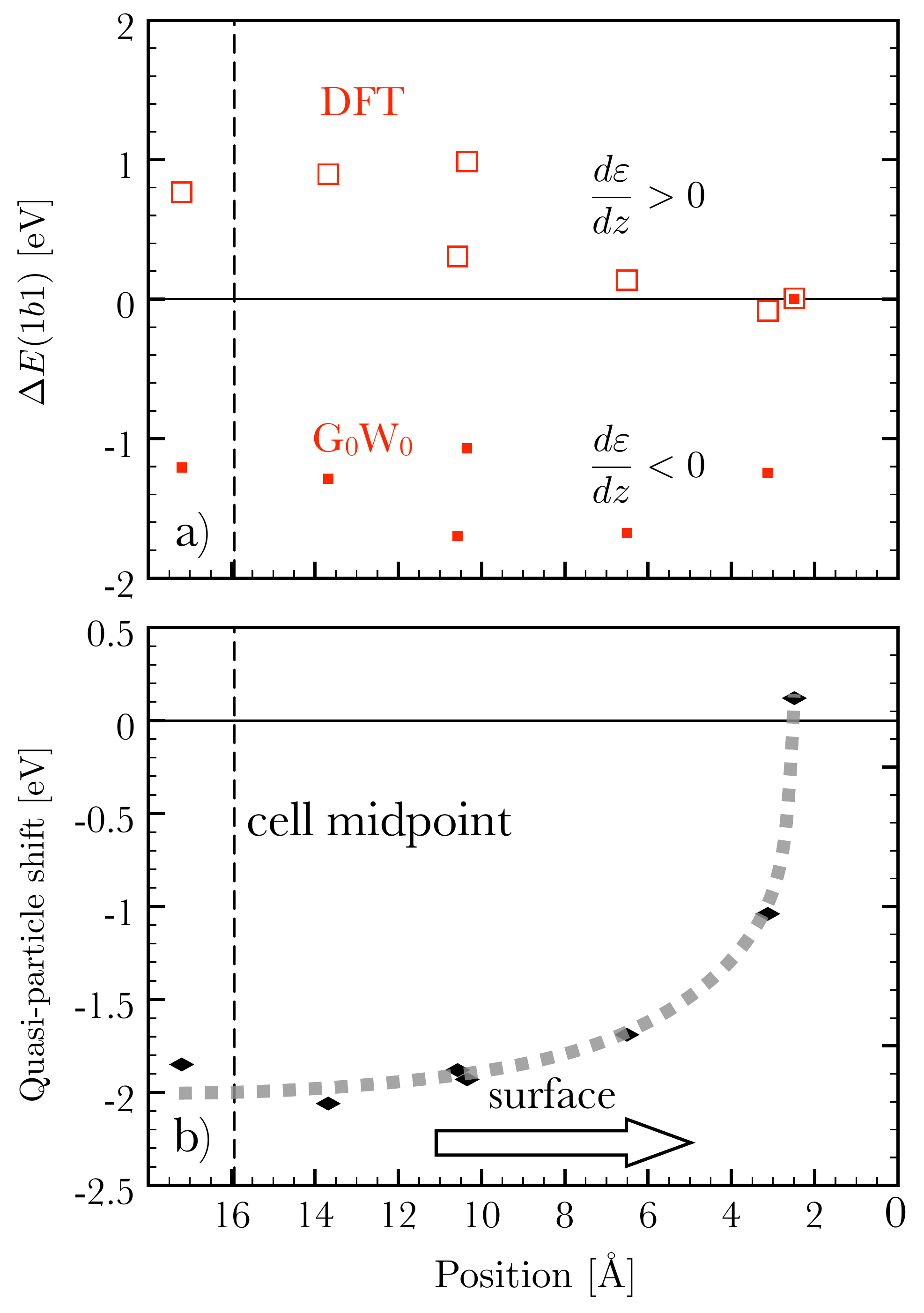}
\end{center}
\caption{\label{f:qp_shift} a) Comparison of $\Delta E(1b1)$ as a function of position within DFT and G$_0$W$_0$. Levels have been shifted so that the two methods are taken relative to the value obtained for the surface-bound molecule. The position dependence of $\Delta E(1b1)$ in DFT is opposite to that of G$_0$W$_0$.
b) Quasi-particle shift ($\varepsilon_{\textrm{GW}} - \varepsilon_{\textrm{DFT}}$) as a function of position for the same 1b1 state (dashed line is a guide to the eye). The magnitude of this shift depends both on the position of the molecule within the liquid and its local bonding environment. The rightmost water molecule shown here is adsorbed to the surface and is strongly coupled to it (see Fig.~\ref{f:matrix}c). A molecule at a similar height above the surface (Fig.~\ref{f:matrix}b), is less coupled to the surface and has a much larger associated quasi-particle shift.}
\end{figure}

For the remaining molecules, several key features are apparent. DFT severely underestimates the position of molecular orbital level alignment compared to experiment (the difference is $\approx$ 3-4 eV). Moreover, the position dependence (with respect to the metal surface) for all three orbitals is very weak. For 1b2, there is essentially no variation (switching to a hybrid functional is unlikely to change this~\cite{biller_jcp_2011}). For 1b1 and 3a1, the slope of $\varepsilon (z)$ is positive; orbitals associated with molecules near the surface are deeper in energy than those farther away (in the bulk). Importantly, DFT predicts $\frac{d\varepsilon}{dz}$ which is both smaller and \emph{of opposite sign} to that of G$_0$W$_0$  (see Fig.~\ref{f:qp_shift}a for comparison). Level alignment predicted within G$_0$W$_0$ is in better agreement with experiment (including the relative spacing the three frontier orbitals) and $\frac{d\varepsilon}{dz}$ is much more physically reasonable. Ensuring the correct behaviour for the position dependence of electron removal energies will be important when applying methodologies designed to consider charge transfer and rates~\cite{pavanello_jcp_2013}, molecular scale transport~\cite{toher_prl_2005, quek_nanolett_2007, changsheng_prb_2009, rangel_prb_2011}, electrochemical responses to external potentials, and other excited state processes~\cite{prezhdo_acr_2009}.

It is important to note that despite the improvement relative to DFT, G$_0$W$_0$ appears not to be in complete agreement with experiment. Although this G$_0$W$_0$ calculation is a significant improvement over the DFT solution, several approximations have been made which could impact the fidelity of the result. As previously mentioned, subsystem errors carry over to the interface and can affect relative alignments. Additionally, spin-orbital coupling (known to impact structure of the d band of metals such as Pt) is not included. As Pt is not a simple metal, treating the frequency dependence of the dielectric function through a plasmon pole model clearly leaves room for improvement (vertex corrections may also be necessary~\cite{mahan_prl_1989}). Indeed, the work function of Pt differs from experiment by 7\%~\cite{singh_prb_2009} (there is however significant spread of experimental values~\cite{nieuwenhuys_surfsci_1976, hulse_appsursci_1980, derry_prb_1989}) Finally, as was noted that for the case of pure water~\cite{pham_prb_2014}, the details of geometry (defined by the \emph{ab initio} method or force field used to describe interactions) itself can contribute to differences in both the absolute level position and gap.

Given the extreme cost of treating systems like this with G$_0$W$_0$, it is clear that approximate methods would be of great use when treating similar systems. One approach may be to adapt the successful ``image-charge'' model developed by Neaton et al.~\cite{neaton_prl_2006}. Note that the water platinum interface is quite complex, however, as it involves a mixture of bonding environments, and has transient species present. The application of an external electric field will surely complicate matters further~\cite{xia_prl_1995, otani_jpsj_2008}. While developing a correction is beyond the scope of this work, it is possible to use the results of this calculation to determine what such a correction would look like.

In Fig.~\ref{f:qp_shift}b, I plot the quasi-particle shift ($\varepsilon_{\textrm{GW}} - \varepsilon_{\textrm{DFT}}$) for several molecules in the liquid. A correction such as this could be applied to ground state DFT calculations using an approach which was originally designed to deal with transition metal oxides (DFT+V$_w$ ~\cite{ivady_arxiv_2014}). Such an approach must however respect the bonding which occurs at the surface. This can likely be accomplished through approaches such as molecule centered sub-matrix diagonalization~\cite{quek_nanolett_2007}, or simply observing the broadening of the oxygen p-orbital character.


In conclusion, I have shown results for the frontier orbital level alignment of liquid water at a Pt(111) surface obtained through many-body perturbation theory in the G$_0$W$_0$ approximation. The interface was generated through \emph{ab initio} molecular dynamics simulations. The relative separation of frontier orbitals (1b1,3a1, and 1b2) is well described within G$_0$W$_0$ for the isolated molecule and the interface. In the basis of gas phase orbitals, the diagonal-approximation holds. Any form of model self-energy correction should depend both on the relative position of molecules with respect to the surface and their degree of interaction with it. Level alignment within DFT is \emph{qualitatively} different than G$_0$W$_0$ for this system, a finding which could have important consequences for predicting and interpreting the chemical and physical processes which occur at this interface. 

\section{Acknowledgments}

Work supported by NSERC, ComputeCanada, and SOSCIP. The author would like to acknowledge fruitful discussions with P. Doak, T. Ogitsu, D. Prendergast, S. Sharifzadeh, D. Strubbe, and S.Y. Quek.

\bibliography{references}

\end{document}